\documentclass[aps,pre,floats,floatfix,preprint,superscriptaddress,showkeys]{revtex4}

\usepackage{latexsym,amsmath}
\usepackage{amsbsy}
\usepackage[pdftex]{graphicx}
\usepackage{rotating}
\usepackage[usenames]{color}
\usepackage{amssymb}
\usepackage{multirow}
\usepackage{float}

\begin{document}

\title{Assessing the significance of knockout cascades in metabolic networks}

\author{Oriol G\"uell}
\thanks{E-mail Corresponding Author: oguell@ub.edu}
\affiliation{Departament de Qu\'{i}mica F\'{i}sica, Universitat de Barcelona, Mart\'{i} i Franqu\`{e}s 1, 08028 Barcelona, Spain}
\author{Francesc Sagu\'es}
\affiliation{Departament de Qu\'{i}mica F\'{i}sica, Universitat de Barcelona, Mart\'{i} i Franqu\`{e}s 1, 08028 Barcelona, Spain}
\author{Georg Basler}
\affiliation{Systems Biology and Mathematical Modeling, Max Planck Institute for Molecular Plant Physiology, Am M\"{u}hlenberg 1, 14476 Potsdam, Germany}
\author{Zoran Nikoloski}
\affiliation{Systems Biology and Mathematical Modeling, Max Planck Institute for Molecular Plant Physiology, Am M\"{u}hlenberg 1, 14476 Potsdam, Germany}
\author{M. \'Angeles Serrano}
\affiliation{Departament de F\'{i}sica Fonamental, Universitat de Barcelona, Mart\'{i} i Franqu\`{e}s 1, 08028 Barcelona, Spain}

\begin{abstract}
Complex networks have been shown to be robust against random structural perturbations, but vulnerable against targeted attacks. Robustness analysis usually simulates the removal of individual or sets of nodes, followed by the assessment of the inflicted damage. For complex metabolic networks, it has been suggested that evolutionary pressure may favor robustness against reaction removal. However, the removal of a reaction and its impact on the network may as well be interpreted as selective regulation of pathway activities, suggesting a tradeoff between the efficiency of regulation and vulnerability. Here, we employ a cascading failure algorithm to simulate the removal of single and pairs of reactions from the metabolic networks of two organisms, and estimate the significance of the results using two different null models: degree preserving and mass-balanced randomization. Our analysis suggests that evolutionary pressure promotes larger cascades of non-viable reactions, and thus favors the ability of efficient metabolic regulation at the expense of robustness.
\end{abstract}

\keywords{metabolic networks; robustness; cascading failure; null models}

\maketitle

\section{Introduction}

Complex networks \cite{Albert:2002,Dorogovtsev:2008,Barrat:2008} may be affected by structural perturbations that modify their behavior or even lead to their collapse. Of all potential structural perturbations, the removal of individual nodes has received most attention. It has been found that complex networks are robust to accidental failures, yet fragile to targeted attacks knocking down their most connected nodes \cite{Havlin:2000,Barabasi:2000b}. A similar approach has been applied in biology to metabolic networks which, as compared to other large-scale cellular networks, prove to be of particular interest due to the availability of high quality reconstructions based on complementary sources of experimental data, and due to the possibility of experimentally validating computational predictions \cite{Szalay:2007,Motter:2008,Serrano:2011b}. Metabolic networks, modeled as complex networks \cite{Palsson:2006,Ma:2003a,Guimera:2005b,Serrano:2012a,Holme:2003}, have been tested in terms of robustness in a wide variety of in silico experiments.

Typical \textit{in silico} strategies to quantify \textit{structural robustness} of metabolic networks against removal of a single reaction consider the downstream effect of the removal and the size of the resulting cascade. Properties of the cascade, e.g., size or length, could then be considered as proxies for the potential damage inflicted by the removal. Finally, one would like to obtain estimates for the statistical significance of the obtained observations, which could be empirically carried out with respect to a well-chosen null model. For instance, by applying this strategy in combination with degree preserving randomization (see Sections 2.2 and 2.3), Smart and coworkers \cite{Smart:2008} have determined that bacterial organisms may have evolved towards reducing the probability of having large cascades, thus, increasing robustness. 

Recently, more complicated perturbations have also been considered, namely, removal of pairs of reactions and sets of genes \cite{Guell:2012}, and the topological significance of the results was evaluated with respect to degree preserving randomization. The employed degree preserving null model has its roots in the Configuration Model \cite{Molloy:1995} for bipartite networks \cite{Newman:2001,Newman:2002,Guillaume:2006}, that  results in randomized network variants in which node degrees are preserved. However, such a degree preserving randomization does not account for the most basic physico-chemical constraints, and may lead, in the case of metabolic networks, to consideration of a reaction which is not mass (i.e., stoichiometrically) balanced (a reaction which does not preserve the same type and number of atoms on its substrate and product sides). As a result, the randomized networks may not be chemically feasible. As an alternative, a novel null model called mass-balanced randomization has been recently proposed \cite{Basler:2011} to account for this issue. Mass-balanced randomization results in randomized network variants in which: (1) every reaction is mass balanced and (2) degrees of reactions are unaltered. By comparing the differences between the analyzed property in the original network and in its randomized variant, this null model is able to distinguish between properties which result solely from physico-chemical constraints and those selected for by evolutionary pressure \cite{Basler:2012}.

Here we explore the metabolic networks of {\it Staphylococcus aureus} \cite{Smart:2008} and {\it Escherichia coli} \cite{Feist:2007} to determine the extent to which the robustness against the failure of individual reactions and pairs of reactions, as quantified by the number of non-viable reactions caused upon the initial removal(s) of a single reaction or a pair of reactions, is bounded by structural constraints. To this end, we compare cascades in the original networks with those obtained from the two null models referred above: degree preserving (DP) and mass-balanced (MB) randomization. We find that the two null models give very different results, which is explained in terms of the properties of the networks obtained by both randomization methods. We use Kolmogorov-Smirnov tests \cite{Smirnov:1948} to statistically assess whether the null models are enough to explain the resulting damage distributions in the original networks. Interestingly, we find that the two organisms exhibit cascades whose properties lie between those expected from the two considered null models, which suggest that factors other than node degrees or physical principles affect the considered features. Moreover, our findings point out that, in the analyzed metabolic networks, evolutionary pressure may not have lead towards minimized damage spreading, which opposes earlier findings based solely on degree preserving randomization. Our results reinforce the importance of choosing an appropriate null model according to the question at hand, since the null model ultimately affects the interpretation of the findings.

\section{Methods}

\subsection{Representation of metabolic networks as complex networks}

We have modeled metabolic networks as bipartite graphs \cite{Holme:2003}, whereby the set of nodes is portioned into metabolites and reactions, and there are no links between nodes of the same type. Metabolites are connected to the reactions by directed links, which allow differentiating between reactants and products. Every reversible reaction is split into two irreversible reaction nodes, so that every reaction node included in the representation corresponds to an irreversible reaction. 

We use two different bacterial organisms in this study. {\it E. coli} is the most studied prokaryotic organism and it is the bacterial model which is most frequently used due to ease of experimental manipulation. To construct the bipartite representation of the metabolic network of {\it E. coli} \cite{Feist:2007}, we use data from the BiGG database\footnote{http://bigg.ucsd.edu/}. The resulting network contains $N_R=2066$ reactions and $N_M=1667$ metabolites. 

{\it S. aureus} is an anaerobic bacterium which is present world-wide. Its bipartite directed network reconstruction was obtained from \cite{Smart:2008}, and consists of $N_R=640$ reactions and $N_M=644$ metabolites. 

The networks of the two analyzed species show marked similarities with respect to the cumulative degree distributions. For metabolites, the degree distribution shows a typical power-law form, $P(k) \sim k^{-\gamma}$, with an exponent $\gamma=2.13$ for {\it S. aureus} and $\gamma=2.09$ for {\it E. coli}. On the contrary, reactions show a peaked distribution centered at the average degree corresponding to each network, having a value of $<k_R>=4.8$ for {\it S. aureus} and  a value of $<k_R>=4.3$ for {\it E. coli}.

\subsection{Cascading failure algorithm}

\begin{figure}
 \centering
 \includegraphics[width=0.57\textwidth]{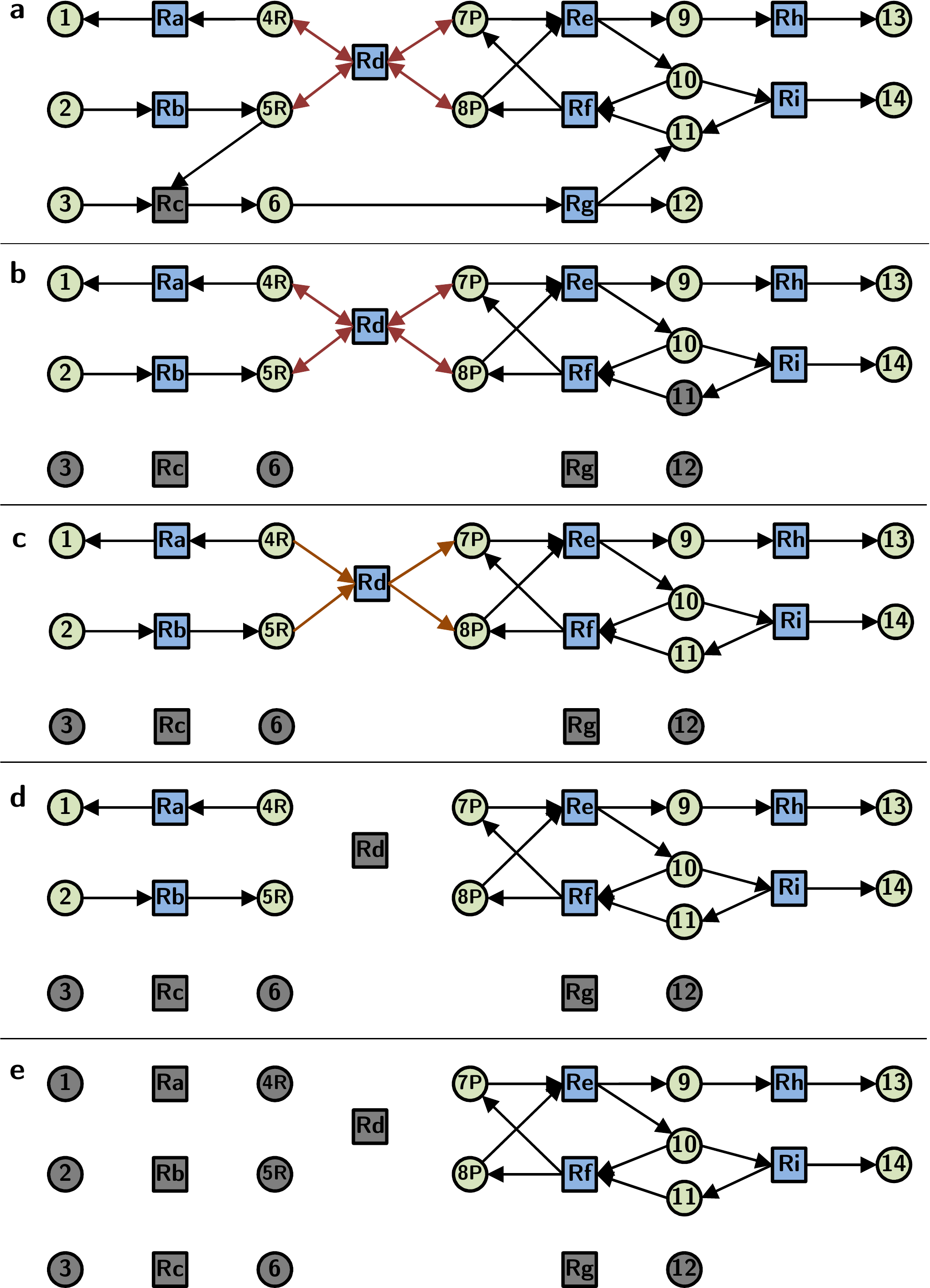}
 \caption{Example of the application of the cascading failure algorithm to a metabolic network. a) For clarity, metabolites 4 and 5 are labeled with $R$ and 7 and 8 with $P$ depending on whether they are reactants or products of the reversible reaction. The cascade starts when reaction $Rc$ fails. b) Therefore, metabolites 3 and 6 become non-viable. Because metabolite 6 is connected to reaction $Rg$, the later becomes non-viable, turning also metabolite 12 non-viable. Notice that metabolite 11 loses one {\it in} connection, but it is still viable, meaning that one of the waves of the cascade stops here. However the other wave keeps spreading. c) Metabolite 5 becomes inviable, causing the reversible reaction $Rd$ to remain viable only towards the production of metabolites 7 and 8. e) Consequently, metabolite 4 becomes non-viable, and so its associated reactions also become non-viable. The cascade spreads until all metabolites and reactions affected by the cascade remain viable. Finally, note that metabolites 1, 2, 3, 13, and 14, which initially have no incoming or outgoing connections, are not considered inviable by the algorithm.}
 \label{fig:1}
\end{figure}

A cascading failure algorithm \cite{Smart:2008} is applied to spread the initial perturbation through the network and to compute the corresponding damage. Crucial to the algorithm are the concepts of viable metabolites and reactions. A metabolite is considered viable if it has at least one incoming and one outgoing connection, so as to prevent depletion or accumulation. This structural condition is a prerequisite for the network as a whole to operate at a positive steady state. On the other hand, reactions are viable if and only if all of the participating metabolites are viable. The algorithm then starts with a network from which an initial set of reactions is removed. In the following step, all reactions and metabolites that, as a consequence, become non-viable are removed, which in its turn results in additional changes of the viability status of the nodes. When only viable reactions and metabolites remain in the network, the damage inflicted by the initial perturbation is quantified as the final number of non-viable reactions (see Figure \ref{fig:1}). Here, we consider perturbations by initially removing every single reaction and each pair of reactions. 

Benchmark cascades are computed by first randomizing the original network and then performing the cascade algorithm on the randomized version. For each of the two null models, damage distributions are obtained for a hundred realizations, then averaged, and finally compared to the original damage distribution to assess the statistical significance and quantitative behavior of the observation.

\subsection{Degree preserving randomization}

The degree preserving randomization method approximates the Configuration Model for bipartite networks \cite{Molloy:1995,Newman:2001,Newman:2002,Guillaume:2006} and works as follows. A pair of links of the network is chosen at random and their targets are swapped, unless this would lead to the repeated occurrence of a metabolite in a reaction. The randomization algorithm is summarized as follows:

\begin{enumerate}
 \item Pick two links at random: $m_1 \rightarrow r_1$ and $m_2 \rightarrow r_2$ or $r_1 \rightarrow m_1$ and $r_2 \rightarrow m_2$.
 \item Swap the end of the links avoiding repeated links and self-production: ($m_1 \rightarrow r_2$ and $m_2 \rightarrow r_1$ or $r_1 \rightarrow m_2$ and $r_2 \rightarrow m_1$).
 \item Repeat until we perform $L^2$ swappings, where $L$ is the total number of links. 
 \item Make several realizations of the randomized metabolic network following the three previous steps.
\end{enumerate}

The links from reversible reactions are rewired independently of those from irreversible reactions, using the same steps previously mentioned, in order to preserve the degrees of metabolites corresponding to reversible and irreversible reactions, respectively. A scheme showing this algorithm is shown in Figure \ref{fig:2}. It is easy to show that the networks obtained using this method preserve the degrees of both metabolites and reactions, as illustrated in Figure \ref{fig:1b}.

\subsection{Mass-balanced randomization}

\begin{figure}
 \centering
 \includegraphics[width=0.82\textwidth]{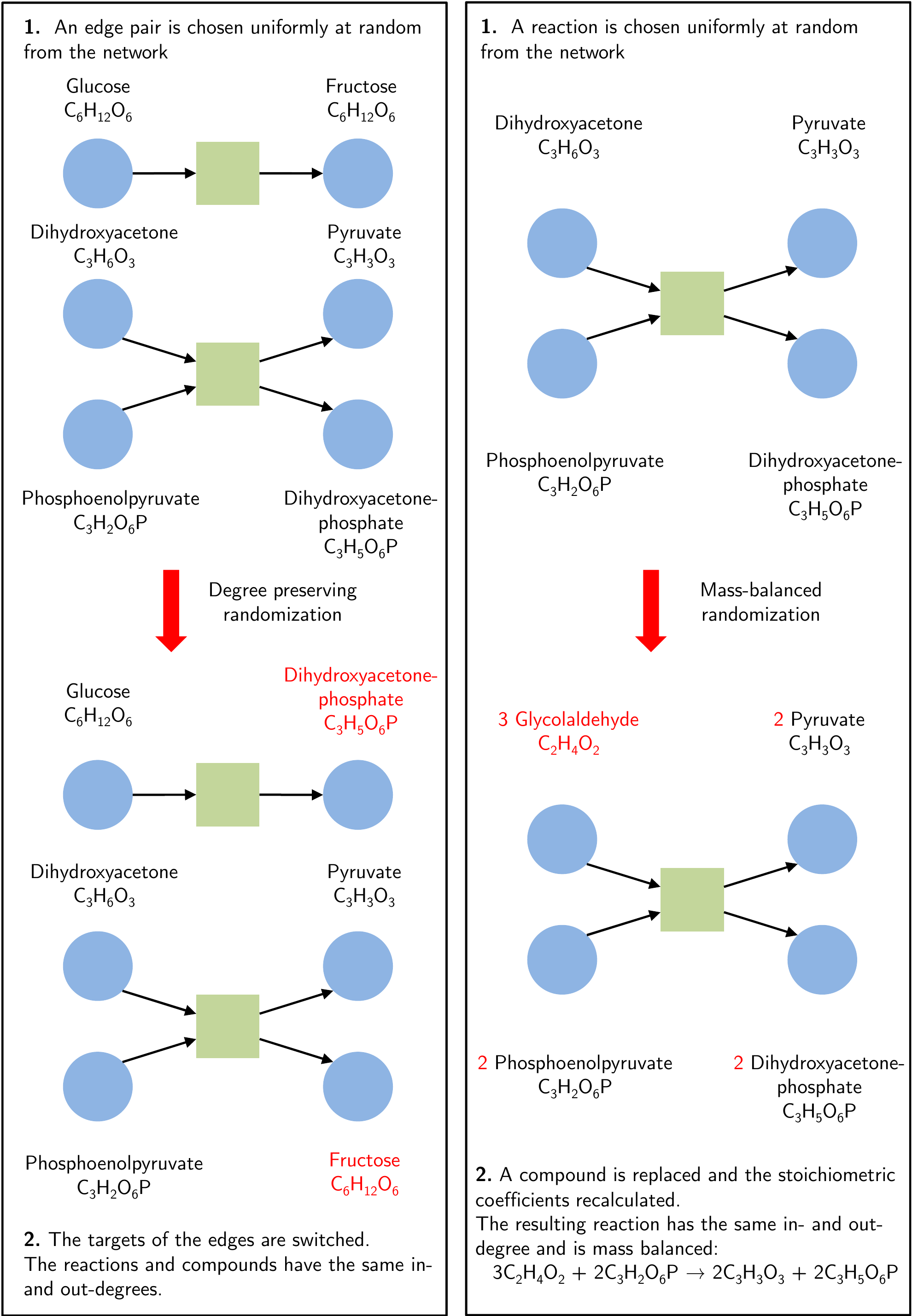}
 \caption{{\it Left}: Scheme of the degree preserving randomization algorithm. In- and out-degrees are conserved, but mass balance is not satisfied. {\it Right}: Scheme of the mass-balanced randomization. In this case metabolites are switched only if the new reaction is mass balanced; while reaction degrees are kept constant, the degrees of metabolites are not preserved.}
 \label{fig:2}
\end{figure}

Mass-balanced randomization generates randomized networks by rewiring the links corresponding to substrate-reaction or product-reaction relationships, while preserving atomic mass balance of the reactions \cite{Basler:2011}. Given a reaction $r$, its atomic mass balance is given by:\begin{equation}\label{eq:mass_balance}
\sum_{e \in E_r}{s_{e,r} \cdot m_e} = \sum_{p \in P_r}{s_{p,r} \cdot m_p},
\end{equation}
where $E_r$ is the set of substrates and $P_r$ the set of products in $r$, $m_e, m_p$ are the vectors of sum formulas of $e$ and $p$, respectively, and $s_{e,r}$, $s_{p,r}$ their stoichiometric coefficients. For instance, consider the reaction $A \rightarrow B$, with $m_{A}=$ $m_{B}=$ $\textnormal{C}_6\textnormal{H}_{12}\textnormal{O}_6$. Then, $A$ may be substituted by a compound $C$ with $m_{C}=\textnormal{C}_3\textnormal{H}_{6}\textnormal{O}_3$ from within the network, resulting in the randomized reaction 2 $C \rightarrow B$, which satisfies Equation \ref{eq:mass_balance} since  $2~\textnormal{C}_3\textnormal{H}_{6}\textnormal{O}_3 = \textnormal{C}_6\textnormal{H}_{12}\textnormal{O}_6$ (Figure \ref{fig:2}). In addition to substituting individual substrates or products, the method also allows more complex substitutions involving pairs of substrates or products, yielding a large number of possible substitutions.

\begin{figure}
 \centering
 \includegraphics[width=0.85\textwidth]{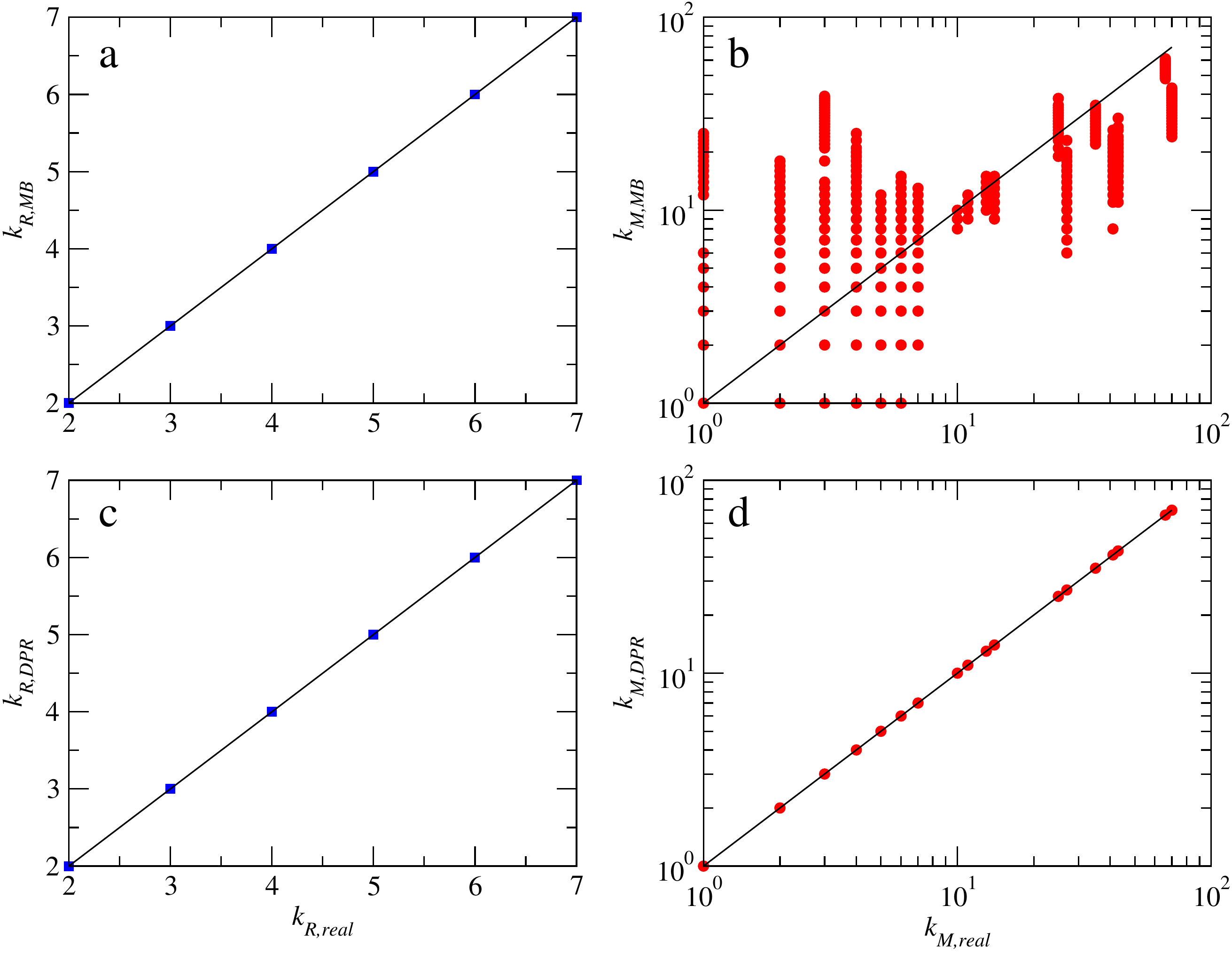}
 \caption{Comparison of the degrees of reactions and metabolites obtained by the two null models applied to the network of {\it E. coli}. In this representation, each point is a reaction or a metabolite with coordinates ($k_{real}$,$k_{randomized}$), where $k_{real}$ is the metabolite/reaction degree in the original network, and $k_{randomized}$ the corresponding degree in a randomized network. Points fall in the diagonal if degrees are preserved in the randomized networks. a) Mass-balanced randomization (MB). This method gives networks in which the degrees of the reactions are preserved. However, b) degrees of metabolites are not conserved. c) Degree preserving randomization (DP). This method gives networks with preserved degrees of reactions. d) Degrees of metabolites are also preserved with degree preserving randomization, however at the expense of violating mass balances of reactions.}
 \label{fig:1b}
\end{figure}

The motivation for preserving atomic mass balance of reactions, a fundamental physico-chemical constraint, is that the resulting null model allows estimating the importance of network properties with respect to evolutionary pressure. As biological systems and their properties evolve under physical constraints and evolutionary pressure, a null model which satisfies physical principles but does not account for evolutionary pressure differs from a metabolic network only in the properties which are affected by evolutionary pressure. Thus, a property deemed statistically significant following mass-balanced randomization is beyond basic physical constraints and likely to be a result of evolutionary pressure \cite{Basler:2012}. Again, a hundred randomized networks are generated from a given metabolic network. The method preserves mass balance and reaction degrees, while the stoichiometric coefficients and metabolite degrees are changed (Figure \ref{fig:1b}a, b).

\section{Results}

\subsection{Individual removal of reactions}

First, we study cascades triggered by individual removal of reactions, from $r=1$ to $r=N_R$. We define the size of a cascade, $d_r$, or damage caused by the removal of a reaction $r$ as the number of resulting non-viable reactions.

\begin{figure}[ht]
 \centering
 \includegraphics[width=0.85\textwidth]{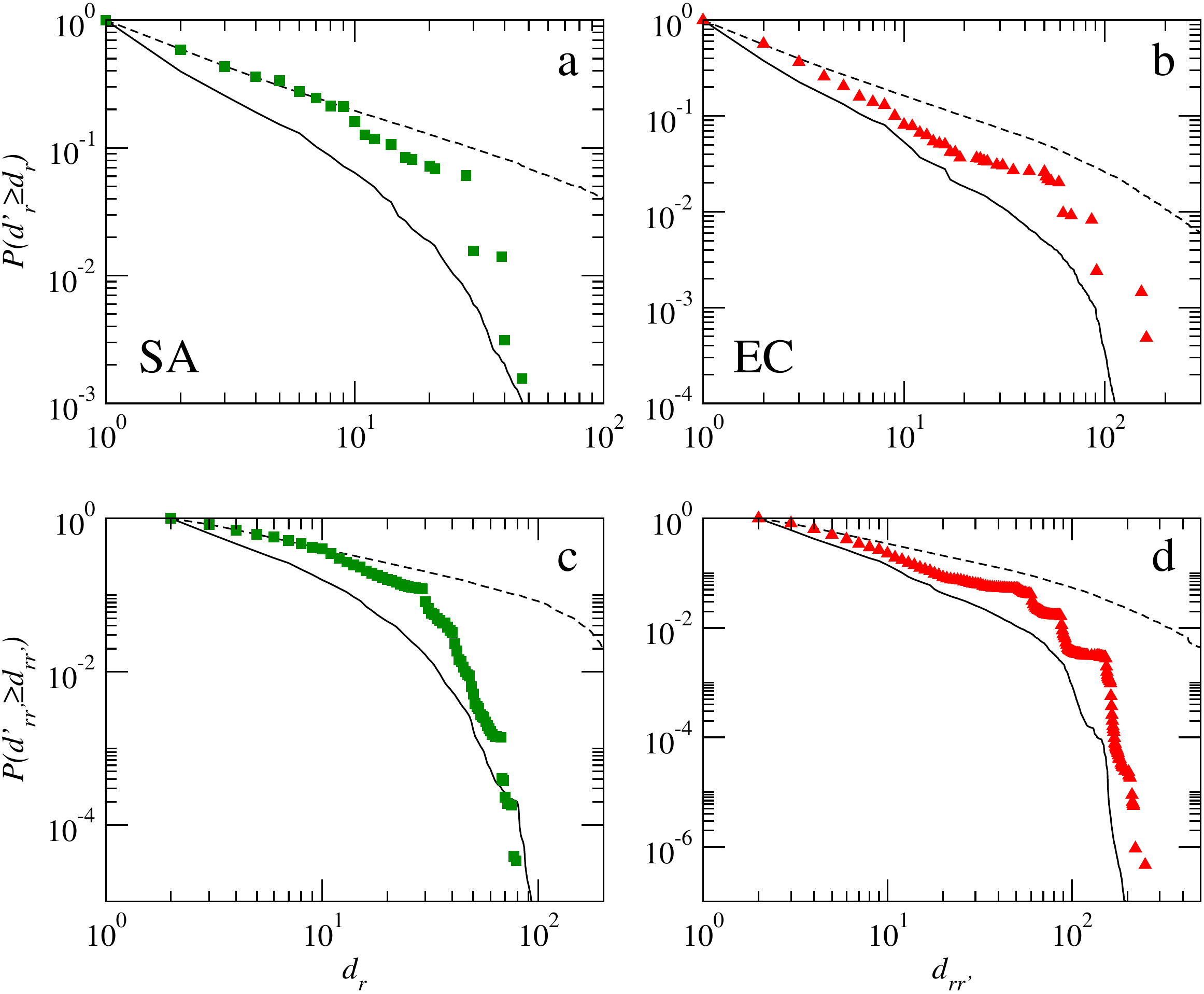}
 \caption{Distributions of damage caused by removal of reactions. a-b) Cumulative probability distributions for {\it S. aureus} (green) and {\it E. coli} (red). Averaged distributions over 100 randomizations of the original networks are shown for degree preserved (dashed line) and mass-balanced randomization (continuous line). c-d) Damages caused by pairs of removal of reactions, with the same symbols as in a-b).}
 \label{fig:3}
\end{figure}

The cumulative probability distributions $P(d'_{r} \geq d_r)$ give the probabilities that a damage is at least as large as $d_r$, and may be interpreted as a measure of the robustness of a network with respect to reaction removal. We determined the cumulative probability distributions of damage in the two metabolic networks under analysis in this study and compared them to the averaged distributions associated to their randomized variants from degree preserving and mass-balanced randomization (Figure \ref{fig:3}). 

All distributions show a similar power-law distribution, with an exponential cut-off indicating possible finite size effects. The detected heterogeneity indicates that the failure of most reactions results in small damage, while some specific reactions affect large parts of the corresponding network, which is in agreement with the robustness previously attributed to metabolic networks \cite{Barabasi:2000b}. However, here we observe that the same trend also holds for the randomized networks, both from degree preserving and mass-balanced randomization. Thus, the heterogeneous behavior of the damage distributions is a general feature which is not specific to metabolic networks, and thus insufficient to explain robustness of metabolism.

Instead, when comparing the distributions of damages between the original and randomized networks,  we observe that the distributions of the original networks lie in between the distributions of the two null models. Particularly, in both organisms, damages are smaller compared to their degree preserving randomizations, but larger when compared to their mass-balanced randomizations. Thus, the robustness of the analyzed networks cannot be explained by the distribution of degrees or by basic physical constraints. For the DP null model, this finding indicates that robustness is positively influenced by factors which are independent of the degrees. The results from the MB null model suggest that evolutionary pressure leads to larger cascades of non-viable reactions, and thus lower robustness (see Discussion).

At first glance, the damage distributions of the original networks seem to closely resemble those of the mass-balanced randomized networks. To check whether or not the cumulative probability distributions significantly differ between the original networks and their randomized variants, we perform Kolmogorov-Smirnov tests \cite{Smirnov:1948} (Table \ref{fig:t1}). The null hypothesis is that the two samples are drawn from the same distribution, which is rejected if the obtained $p$-value is no greater than the standard significance level $\alpha = 0.05$. In this case, the compared distributions are considered significantly different.
\begin{table}
 \centering
 \begin{tabular} {|c|c|c|c|c|}
  \hline
  \multirow{2}{*}{Organism}&\multicolumn{2}{c|}{SR}&\multicolumn{2}{c|}{PR}\\
  \cline{2-5}
  &MB&DP&MB&DP\\
  \hline
  {\it S. aureus}&$0.19$/$0$&$0.085$/$0.0002$&$0.27$/$0$&$0.15$/$0$\\
  {\it E. coli}&$0.19$/$0$&$0.082$/$2 \cdot 10^{-12}$&$0.21$/$0$&$0.13$/$0$\\
  \hline
 \end{tabular}
 \caption{Kolmogorov-Smirnov tests for comparing single reaction (SR) and pairs of reactions (PR) failure cascades in the two metabolic networks with both randomization methods, mass-balanced (MB) and degree preserving (DP). We give the values of the KS statistic / associated significance level.}
 \label{fig:t1}
\end{table}

\subsection{Failure of pairs of reactions}

In the following, we analyze failure cascades resulting from the removal of each possible pair of reactions, where both $r$ and $r'$ run from $r,r'=1$ to $r,r'=N_R$ ($r \ne r'$). Similar to the previous section, we determine the cumulative probability distributions $P(d_{rr'}' \geq d_{rr'})$ of the damage $d_{rr'}$ resulting from the knockout of two reactions $r$ and $r'$, which provides a measure for the robustness of a network with respect to the failure of reaction pairs (Figure \ref{fig:3}).

Both organisms {\it E. coli} and {\it S. aureus} display again similar results: the distributions of damages in the real networks show a power-law behavior followed by an exponential cut-off. However, in this case the transition is more accentuated. The distributions of the original networks lie again between the distributions of the two null models. Consequently, our observations from Section 3.1 also hold for the failure of reaction pairs: robustness is positively influenced by factors independent of the degrees, but negatively influenced by evolutionary pressure.

Kolmogorov-Smirnov tests were applied in order to test if the distributions given by the null models differ from those in the original network (Table \ref{fig:t1}). As for individual removal of reactions, we chose a threshold value of $\alpha = 0.05$. Again, the distributions of the original networks are significantly different from those of both randomization methods, with $p$-values of $0$.

\section{Discussion}

The obtained results reveal important findings by comparing the damage caused by the failure of single and pairs of reactions in the original metabolic networks to those measured in the randomized variants.

The cascade algorithm produces larger damages in the original networks as compared to those in mass-balanced randomized networks, but smaller cascades as compared to those in degree preserving randomized counterparts. A possible explanation is offered by the difference in global properties of the networks obtained from the two randomization methods \cite{Basler:2012}. Degree preserving randomization decreases the average path length and increases the clustering coefficient of the randomized network, increasing its 'small-world' property. Consequently, such networks are more interconnected, and thus a cascade may in principle propagate further in the network. The opposite holds for mass-balanced randomization, which increases the average path length while decreasing the clustering coefficient of the randomized network, so that the spread of the damage is less likely. Although the average path length does not resemble the length of metabolic inter-conversion, the small-world property may still affect the impact of removal of reactions due to its functional importance. 

In Tables \ref{fig:t1}, the significance levels obtained by the Kolmogorov-Smirnov test show that all distributions obtained from the null models are significantly different from those observed in the original networks. The distributions, though visually similar to the original networks, differ significantly between the original and randomized networks. The fact that the damage distributions of the original networks differ significantly from those in the degree preserved randomized versions allow us to conclude that other properties beyond the degrees are important for determining the size of cascades in the original networks.

We also point out that the principle of cascade propagation relies on violation of a structural precondition for a steady-state, namely that all metabolites can be produced and consumed in order to avoid their depletion or accumulation. However, the steady-state assumption is only meaningful for networks which satisfy fundamental physical principles. We therefore also employed mass-balanced randomization, which guarantees preservation of mass-balance, and thus allows us to discern whether the measured property is a result of basic physical principles, or, instead, whether it is affected by evolutionary pressure. Since the size of cascades in mass-balanced randomized networks is significantly lower than those in real networks, evolutionary pressure may indeed lead to larger cascades.

Consequently, this finding indicates that evolutionary pressure may favor lower robustness of metabolic networks with respect to the failure of reactions, seemingly contradicting the general requirement of robustness in biological systems. However, on one hand, this finding may be a result of the evolutionary versatility of metabolic networks, which favors organisms that are able to evolve quickly, i.e., by few modifications to their metabolic networks. On the other hand, we point out that a cascade may not only be interpreted as the harmful spreading of failure, but also as the ability to regulate metabolism by activating/deactivating reactions, e.g., by transcriptional regulation \cite{DeRisi:1997}. Thus, large cascades, favored by evolutionary pressure, may point at the evolutionary requirement of regulating large parts of metabolism, such as pathways, through the regulation of small sets of enzyme-coding genes. The ability to regulate the activity of metabolic reactions by deactivating competing reactions is a well-known principle of metabolism. Our results thus indicate that evolutionary pressure favors the ability of efficient metabolic regulation at the expense of robustness to gene knockouts, pointing at the necessary integration of trade-offs from various cellular functions.

\begin{acknowledgments}
This work is supported by MICINN Projects No. FIS2010-21924-C02-01 and BFU2010-21847-C02-02; Generalitat de Catalunya grant No. 2009SGR1055; the Ram\'{o}n y Cajal program of the Spanish Ministry of Science, the FPU grant of the Spanish Ministry of Science, and the Max Planck Society.
\end{acknowledgments}

\end{document}